\documentclass[iicol, pdflatex, sn-basic]{sn-jnl}

\usepackage{lineno}
\usepackage{graphicx}
\usepackage{multirow}%
\usepackage{amsmath,amssymb,amsfonts}%
\usepackage{amsthm}%
\usepackage{mathrsfs}%
\usepackage[title]{appendix}%
\usepackage{xcolor}%
\usepackage{textcomp}%
\usepackage{manyfoot}%
\usepackage{booktabs}%
\usepackage{algorithm}%
\usepackage{algorithmicx}%
\usepackage{algpseudocode}%
\usepackage{listings}%

\begin{document}

\title{Analyzing Stellar and Interstellar Contributions to
Polarization: Modeling Approaches for Hot Stars}

\author*[1]{Richard Ignace}
\affil[1]{Department of Physics \& Astronomy,
East Tennessee State University,
Johnson City, TN 37614, USA}

\author[2]{Andrew G. Fullard}
\affil[2]{Institute for Cyber-Enabled Research, Michigan State University, East Lansing, MI 48824, USA}

\author[3]{Georgia V. Panopoulou}
\affil[3]{Department of Space, Earth and Environment, Chalmers University of Technology, 412 93, G\"{o}teborg, Sweden}

\author[4]{D. John Hillier}
\affil[4]{Department of Physics \& Astronomy, University of Pittsburgh, Pittsburg, PA 15260, USA}

\author[1,5]{Christiana Erba}
\affil[5]{Space Telescope Science Institute, 3700 San Martin Drive, Baltimore, MD 21218, USA}

\author[6]{Paul A. Scowen}
\affil[6]{NASA Goddard Space Flight Center, Exoplanets and Stellar Astrophysics Lab, Greenbelt, MD 20771, USA}

\abstract{
Linear polarimetry of unresolved stars is a powerful method for discerning or constraining the geometry of a source and its environment, since spherical sources produce no net polarization. However, a general challenge to interpreting intrinsic stellar polarization is the contribution to the signal by interstellar polarization (ISP). Here, we review methodologies for distinguishing the stellar signal from the interstellar contribution in the context of massive stars. We first characterize ISP with distance using a recent compilation of starlight polarization catalogs. Several scenarios involving Thomson scattering, rapidly rotating stars, optically thick winds, and interacting binaries are considered specifically to contrast the wavelength-dependent effects of ISP in the ultraviolet versus optical bands. ISP is recognizable in the stellar polarization from Thomson scattering in the polarization position angle rotations. For hot stars with near-critical rotation rates, the ISP declines whereas the stellar continuum polarization sharply increases.  In the case of quite dense winds, strong ultraviolet lines trace the ISP, which is not always the case in the optical.  In the binary case, temporal and chromatic effects illustrate how the ISP displaces variable polarization with wavelength. This study clarifies the impacts of ISP in relation to new ultraviolet spectropolarimetry efforts such as {\em Polstar} and {\em Pollux}.  
}


\maketitle

\section{Introduction}

Ground-based visible-band polarimetry has been a mainstay for astrophysical studies from planets to galaxies \citep[e.g.,][]{2006A&G....47c..31H}. The Halfwave spectropolarimeter (HPOL) program provided a wealth of optical data for a variety of stellar sources \citep[e.g.,][]{2012AIPC.1429..226M, 2014JAI.....350009D}. However, there have been only two dedicated space-borne spectropolarimetry missions. The first was the {\em Wisconson Ultraviolet Photo-Polarimetry Experiment} (WUPPE) that focused on low spectral resolution UV polarization observations of planets to galaxies but with an emphasis on stars \citep[e.g.,][]{1994SPIE.2010....2N}. The most recent has been the {\em Imaging X-Ray Polarimetry Explorer} \citep[IXPE;][]{2022JATIS...8b6002W} with a focus on stellar compact objects and active galaxies.  

The polarization of light provides a window into the geometry and magnetic fields of astrophysical sources, even if they are spatially unresolved. This is achieved through the measurement of Stokes parameters I for the total light, Q and U for linear polarization, and V for circular polarization \citep[e.g.,][]{2010stpo.book.....C}.
Hot, massive stars are particularly well suited to investigations using linear polarimetry due to their fast, dense, ionized winds which provide fertile regions for Thomson scattering \citep{2010stpo.book.....C}. Although these stars are rare, their strong winds and explosive ends as supernovae power the formation of new stars and enrich the interstellar medium (ISM) with metals \citep{1995ApJS..101..181W, 2012ARA&A..50..107L}. Understanding their mass loss, including influences of rotation and binarity, is key for interpreting stellar and galactic evolution.

Deviation from spherical geometry in a scattering region can be detected using linear polarimetry, which is the focus of this contribution. Such aspherical scenarios relevant to massive stars include Be stars with discs \citep[e.g.,][]{2013A&ARv..21...69R}, close binary systems with interacting winds \citep[the majority of massive stars are located in binary systems, e.g.,][]{2012Sci...337..444S}, and stochastic clumping in stellar winds \citep[e.g.,][]{2006A&A...454..625P}. 

However, in the UVOIR\footnote{UVOIR refers to the Ultraviolet (UV), optical (O), and Infrared (IR) wavebands of the electromagnetic spectrum.} regime, the majority of linear polarization for Galactic sources derives from the alignment of dust grains in the ISM, giving rise to interstellar polarization (ISP).  In the Milky Way, the wavelength-dependence of ISP is commonly described using the Serkowski Law \citep{1975ApJ...196..261S}, an empirical relation between wavelength and polarization amplitude. The ISP along a sightline to a star is constant with time. That includes the position angle (PA) of the ISP. However, surveys have demonstrated significant variation in the properties of the ISP with both direction and distance \citep{2000AJ....119..923H, 2002ApJ...564..762F, 2014A&A...561A..24B,Siebenmorgen2018}.

While various methods have been used to subtract ISP from polarimetric measurements \citep[for a summary, see][]{2018MNRAS.479.4535S}, it is not always possible to use these methods due to observational constraints. It is therefore important to review modeling strategies that can be employed to disentangle the ISP from the intrinsic polarization. Such strategies are predicated on three key assumptions: (a) the Serkowski Law for ISP with wavelength is valid; (b) the PA for the ISP is constant with wavelength\footnote{However, this may not be the case if there are multiple ISM components along the line of sight with different magnetic field orientations \citep{1974ApJ...187..461M}.}; and (c) both the degree of the polarization and PA for the interstellar contribution are constant with time. 

\begin{figure}[t]
\centering
\includegraphics[width=2.9in]{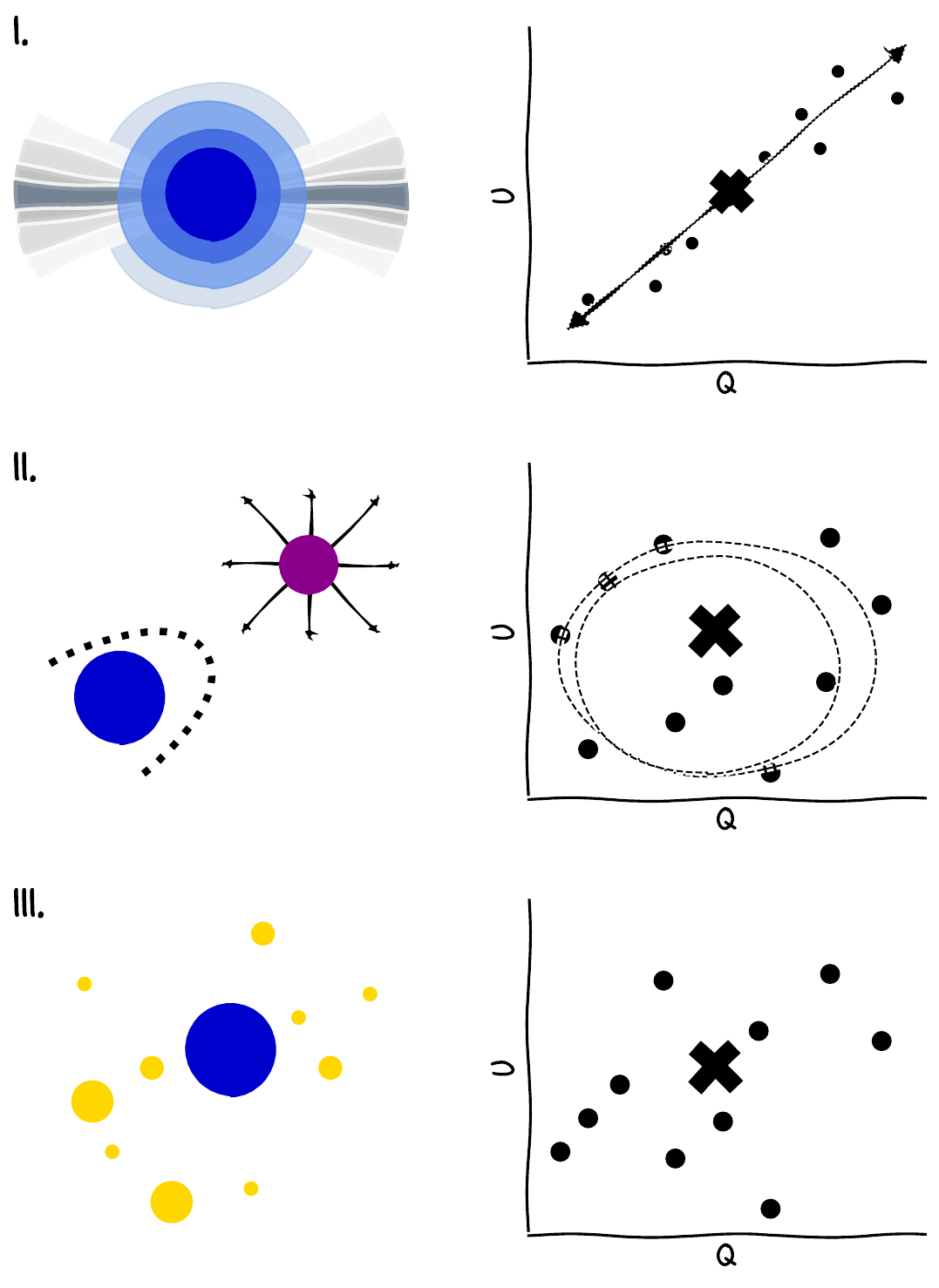}
\caption{Illustrations for how time-variable linear polarization appears in $Q-U$ diagrams for common scenarios. In each diagram, the ``X'' signifies the ISP value. The top row illustrates the ``stationary'' case, as represented by a star with circumstellar disk in which the PA is fixed but the polarization amplitude varies. This produces a linear variation. The middle row illustrates the ``loop'' case. The example is for a binary orbit as one possibility for a secular change in PA on the sky. This maps to loops in the $Q-U$ diagram. The lower row is the ``random'' case, with yellow dots indicating clumps as an example of stochastic structure. Variations in $Q$ and $U$ are randomly distributed about the ISP value.}
\label{figV}
\end{figure}

For context Figure~\ref{figV} provides a cartoon that focuses on intrinsic stellar polarization that is time-varying \citep[based on Figure~3 of ][]{2006ASPC..355..173D}. From top to bottom are scenarios with a fixed axis of symmetry, a changing axis, and the absence of an axis. We denote these categories as I, II, and III, respectively. 

The first (I) uses a circumstellar disk as an example. The disk is axisymmetric. Variable polarization may arise from changing optical depth, or variation in the disk structure (e.g., opening angle) as long as the symmetry axis remains fixed as projected onto the sky. In this case polarization varies along a line in the $Q-U$ diagram. This would be true at every wavelength, modulo absorptive opacity effects. The ISP is signified by the ``X'' and represents a vector translation of the stellar contribution from the origin.

The second (II) scenario makes use of a binary for illustration, but is valid for any situation in which the relevant axis defining the polarization is changing as projected onto the sky \citep[even microlensing; c.f.,][]{1995A&A...293L..46S, 2002MNRAS.336..501S}. For the binary scenario, it is the line of centers joining the two stars that evolves in projection on the sky. In this example, loops are formed as opposed to a line. Once again, an ``X'' marks a fictitious ISP value that shifts the variations away from the origin.

The third (III) scenario is the case of no preferred axis. A clumpy wind is imagined as an example. The result is variable polarization that appears as a random distribution, not only of points for the measurements, but also in terms of their sequencing. In a stochastic process, with sufficient sampling, the centroid of the distribution would correspond to the ``X'' location for the ISM polarization. 

While it is true that time variable polarization can safely be interpreted as intrinsic to the star, the absolute intrinsic polarization remains generally unknown unless ISP is removed.  In a true stochastic process like category III, the centroid is a valid way of determining the ISP.  However, for category II, note that the ISP is not at the center of loops.  And in category I, the location of the ISP along the line is unknown.  Consequently, significant science gains about stellar sources can be made without knowledge of the ISP, yet there remain limitations.

In this paper, we focus on modeling the effects of ISP on observations of hot stars, so that observers can infer stellar properties from observations that have been contaminated by ISP, or even subtract ISP entirely when possible. The methods described will be relevant across the UVOIR range, but especially in the UV where hot stars are brightest and where the {\em Polstar} satellite telescope or the {\em Pollux} instrument would conduct its observations \citep{2022Ap&SS.367..126S, 2024arXiv240915714I, 2024arXiv241001491M}.  

In Section \ref{sec:isp}, current knowledge of the ISP is summarized with reference to survey data. In Section \ref{sec:obs_examples}, observational examples related to Figure~\ref{figV} are given, specifically for categories I and III.  We also expand on the theme by introducing approaches for selected chromatic effects that can be used to determine or constrain the ISP. In Section \ref{sec:infer_intrinsic_pol}, we introduce an overview of modeling approaches relevant to hot stars for inferring the intrinsic stellar polarization. In Section \ref{sec:hybrid_seds}, we detail four specific examples of synthetic ``hybrid'' polarized spectra that combine a Serkowski Law for the ISP with (a) circumstellar electron scattering, (b) a rotationally distorted star, (c) an axisymmetric and multiple scattering wind, and (d) a binary scenario for time variable polarization.  Section \ref{sec:conclusions} provides concluding remarks. 

\section{Interstellar Polarization}\label{sec:isp}

Interstellar polarization arises from the alignment of dust grains with the ambient magnetic field \citep{Davis1951, Lazarian2007, 2015ARA&A..53..501A}. The efficiency of alignment depends on the optical properties of grains, their size distribution, as well as the radiation field. The combination leads to the observed wavelength dependence of the ISP that is well-described by the Serkowski Law \citep{1975ApJ...196..261S}, expressed as
\begin{equation}
  p_\lambda = p_{\rm max}\,e^{-K\,\ln^2(\lambda_{\rm max}/\lambda)},
  \label{eq:serk}
\end{equation}
\noindent where $p_{\rm max}$ is the maximum polarization occurring at wavelength $\lambda_{\rm max}$, and $K$ describes the width of the function. 

\begin{figure}[t]
\centering
\includegraphics[width=3.in]{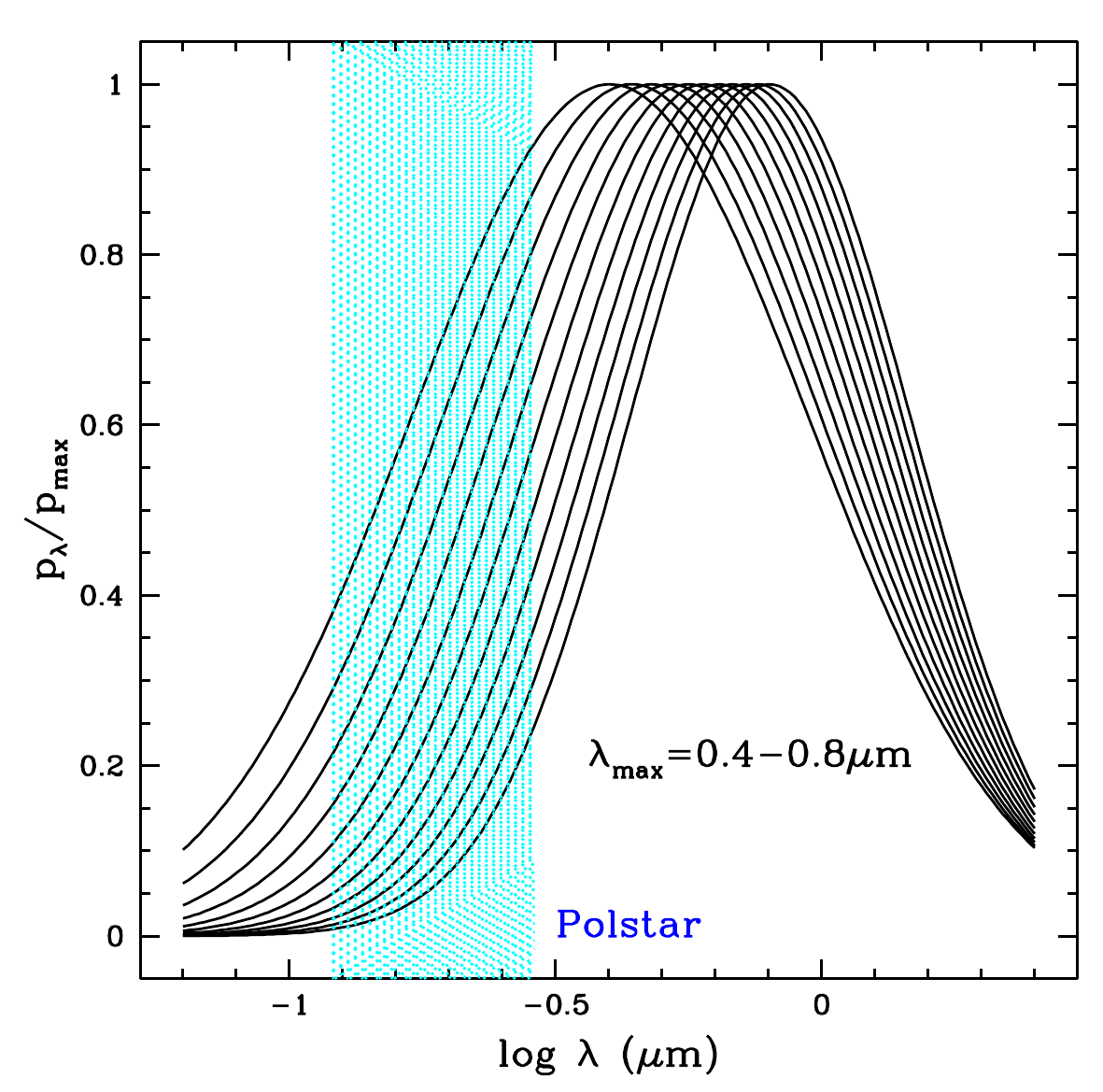}
\caption{A sequence of normalized Serkowski Laws (see eq.~[\ref{eq:serk}]) $p_\lambda/p_{\rm max}$ with logarithmic wavelength. The curves are for a range of $\lambda_{\rm max}$ values as indicated, using $K=1.68\,\lambda_{\rm max}$. The light blue shading indicates the waveband of {\em Polstar}.}
\label{fig6}
\end{figure}

The Serkowski Law typically peaks in the optical and declines toward the UV and the IR. \cite{1992ApJ...386..562W} find $\lambda_{\rm max}$ values ranging from about 400~nm to 800~nm, although values outside this range can sometimes be found. Figure~\ref{fig6} shows example ISP curves using the Serkowski Law for a range of $\lambda_{\rm max}$. The curves are plotted as normalized to $p_{\rm max}$, from the FUV to IR. Overplotted is the waveband for {\em Polstar} in light blue. \cite{1980ApJ...235..905W} find a scaling between the value of $K$ and $\lambda_{\rm max}$. 

Although $\lambda_{\rm max}$ is typically around 0.5 $\mu m$, values of $p_{\rm max}$ vary dramatically between sightlines. Apart from the intrinsic variation of the properties of grains, external factors of the ISM also influence the observed values of the parameters in equation~\ref{eq:serk}. Variations in the interstellar radiation field, which is responsible for aligning the grains, as well as in the gas properties, result in differing efficiencies for alignment as a function of environment \citep{Andersson2010, Voshchinnikov2016, Medan2019,Singh2022, Tram2022}. The geometry of the interstellar magnetic field and the amount of extinction vary with location in the Galaxy, giving rise to additional directional dependence in the observed ISP \citep{Siebenmorgen2018, Angarita2023}. Complexities of the interstellar dust and magnetic field properties may give rise to departures from the Serkowski relation \citep{Mandarakas2024arXiv240910317M}. 

\begin{figure}[t]
\centering
\includegraphics[width=3.in]{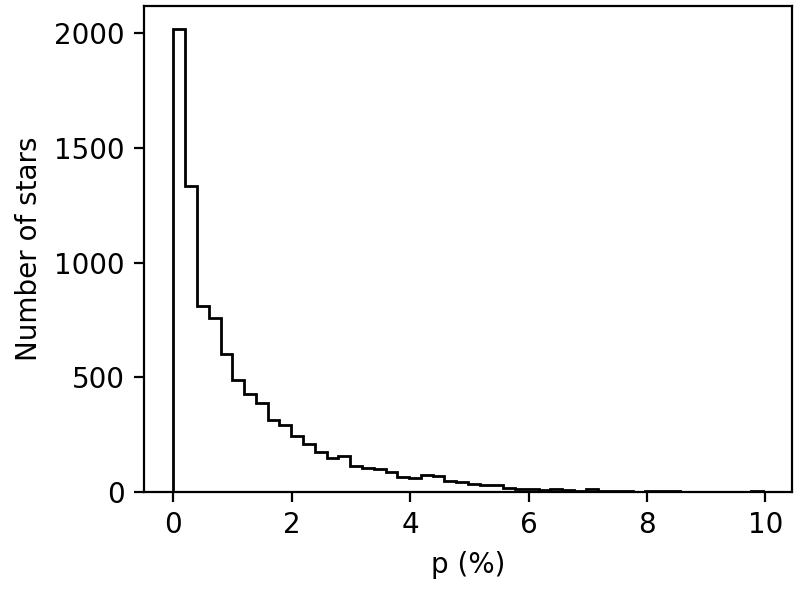}
\caption{A histogram of the incidence of $p$ values for stars within 1~kpc, from the optical polarization catalog of \cite{2025ApJS..276...15P}.}
\label{fig2}
\end{figure} 

We wish to obtain an estimate of the typical ISM $p_{max}$ in the optical for likely targets of UV spectropolarimetry with {\em Polstar} \citep[e.g.,][]{2024arXiv240915714I} using the recent compilation of stellar polarization catalogs by \cite{2025ApJS..276...15P}.  The compilation provides optical polarimetry for $\sim 60,000$ sources and distances from \textit{Gaia}. Only sources with significant detections were used:  adopting a cut at $p/\sigma_p \geq 3$ yields 28,000 entries of high significance. 

Figure~\ref{fig2} shows results for sources within 1~kpc of the Sun, revealing a wide range of polarization fractions of up to 10\%. Note that while the catalog may contain intrinsically polarized sources, the majority of sources are likely not intrinsically polarized.  
\begin{figure}[t]
\centering
\includegraphics[trim={0 2em 0 0},clip, width=3.in]{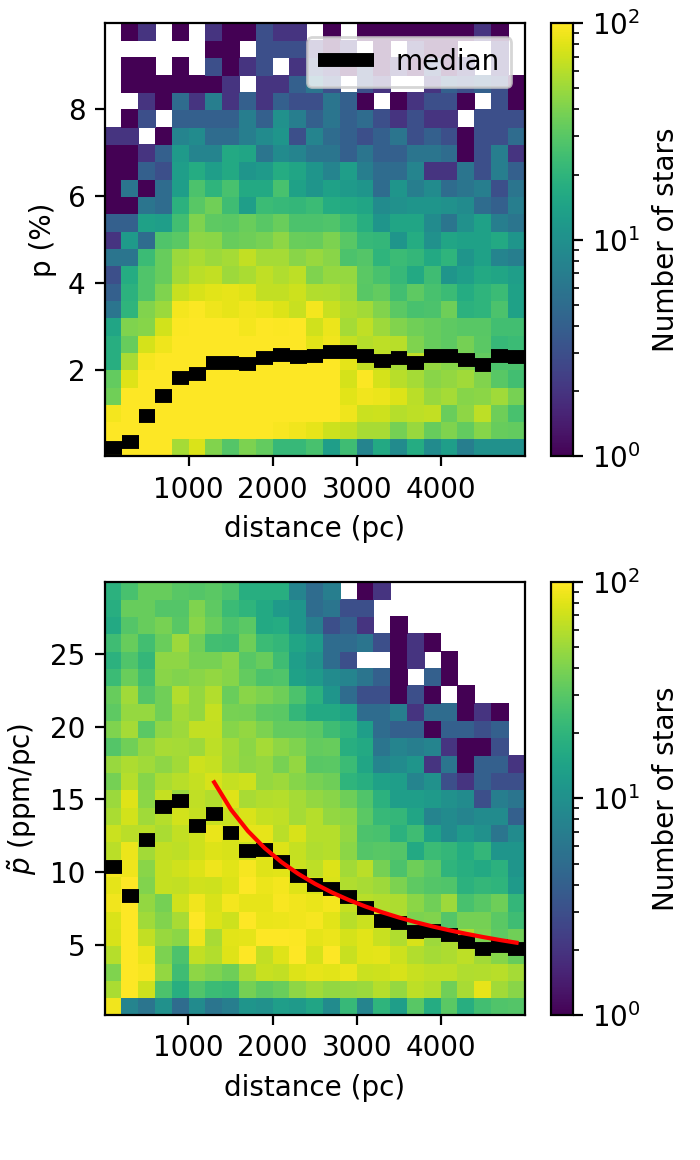}
\caption{Top: A 2D histogram of polarization fraction versus distance for the 28,000 sources with $p/\sigma_{\rm p}\geq 3$ and distances $<$ 5000 pc. Black dots mark the median polarization fraction in bins of distance. Bottom: 2D histogram of $\tilde{p}$ versus distance. Black dots mark the median $\tilde{p}$ per distance bin. A red line shows the best-fit scaling of $d^{-0.9}$.}
\label{fig3}
\end{figure}
The median polarization fraction of the selected sources is 0.7\%. This gives a first estimate for a typical value of ISP within 1~kpc. As these are optical measurements, this estimate approximately corresponds to $p_{\rm max}$ in equation~\ref{eq:serk}.

At farther distances there is a complex interplay between the increase of extinction (which tends to increase the polarization fraction) and the effect of cancellation owing to differently oriented polarization angles from variations in the geometry of the magnetic field. Figure~\ref{fig3} (top) shows the binned distribution of polarization with distance for all sources with $p/\sigma_{\rm p}\geq 3$ and distances $<$ 5000 pc. The median polarization fraction increases up to $\sim$ 1 kpc and then flattens out for larger distances. This behavior is well-known from previous studies with smaller stellar samples \citep[e.g.,][]{2002ApJ...564..762F}. The flattening is a result of two factors: (a) the extinction remains flat beyond 1~kpc for stars at intermediate and high Galactic latitude; and (b) for low Galactic latitude, the effect of polarimetric cancellation comes into play. Note that owing to the inhomogeneous nature of the compilation, there are selection effects which are difficult to quantify; however, the basic trend shown in Figure~\ref{fig3} is expected. 

An alternative way of quantifying the properties of the ISP is by defining a rate of growth \citep[polarization fraction per distance; e.g.,][]{2019MNRAS.483.3636C, 2023AJ....165...87V}:

\begin{equation}
  \tilde{p} = p / d.
\end{equation}

\noindent Here the polarization per distance is expressed as fractional in parts per million (ppm) per parsec. Figure~\ref{fig3} (bottom) shows the 2D histogram of $\tilde{p}$ with distance. Within each distance bin, a median of $\tilde{p}$ is determined and shown as the darkest points. From $1.3$~kpc onwards, there is a decline in the median $\tilde{p}$. We fit the data for bins with $d >1.3$~kpc and find a best-fit scaling of $d^{-0.9}$. 

This drop likely occurs from the effects of polarimetric cancellation.  At large distances, starlight can pass through multiple components with different magnetic field orientations and thus different grain alignments. For uncorrelated orientations the $q$ and $u$ measures will average to zero over increasing numbers of components, thus $\langle q \rangle \approx \langle u \rangle \approx 0$. On the other hand, polarization is positive definite. It does not average to zero; instead, it actually scales roughly as a dispersion with the number of components encountered. If the number of components, $N$, were to increase linearly with distance, $d$, then we would expect $p \propto N^{1/2} \propto d^{1/2}$ \citep{1974ApJ...187..461M}, and $\langle \tilde{p} \rangle \propto d^{-1/2}$. The observed distribution is steeper with an exponent of $-0.9$. But this may arise from selection effects for that data appearing in the catalog. Nonetheless, the expectation of (a) power-law behavior and (b) declining distribution in $\tilde{p}$ is well-motivated.

While these considerations may not apply to any given target, general expectations are valuable for survey planning.
In the following section, we will use typical values for the parameters of the Serkowski relation to explore the effect of ISP in observations of massive stars. A value of $\sim 0.5\%$ for $p_{\rm max}$ will be adopted unless otherwise noted.  This value is consistent with distribution of observed $p$ for stars within 1~kpc. For purposes of illustration, $\lambda_{\rm max}=5500~\AA$ is adopted which corresponds to $K=0.9$ from the scaling $K \approx 1.68~\lambda_{\rm max}~(\mu$m) \citep{1980ApJ...235..905W}. 

\section{Observational Examples}\label{sec:obs_examples}

While there exists a substantial dataset to probe the ISP out to several kiloparsecs, and while the Serkowski Law appears a fairly reliable
prescription for it, removing that signal remains challenging. The Serkowski Law formally involves three independent parameters: $p_{\rm max}$, $\lambda_{\rm max}$, and $K$. In relation to a target star, it is not always the case that field stars in reasonable proximity, both in distance and in direction, yield consistent polarizations in amplitude or PA from which to infer those three parameters.

\begin{figure}[t]
\centering
%[trim={left bottom right top},clip]
\includegraphics[width=3.in]{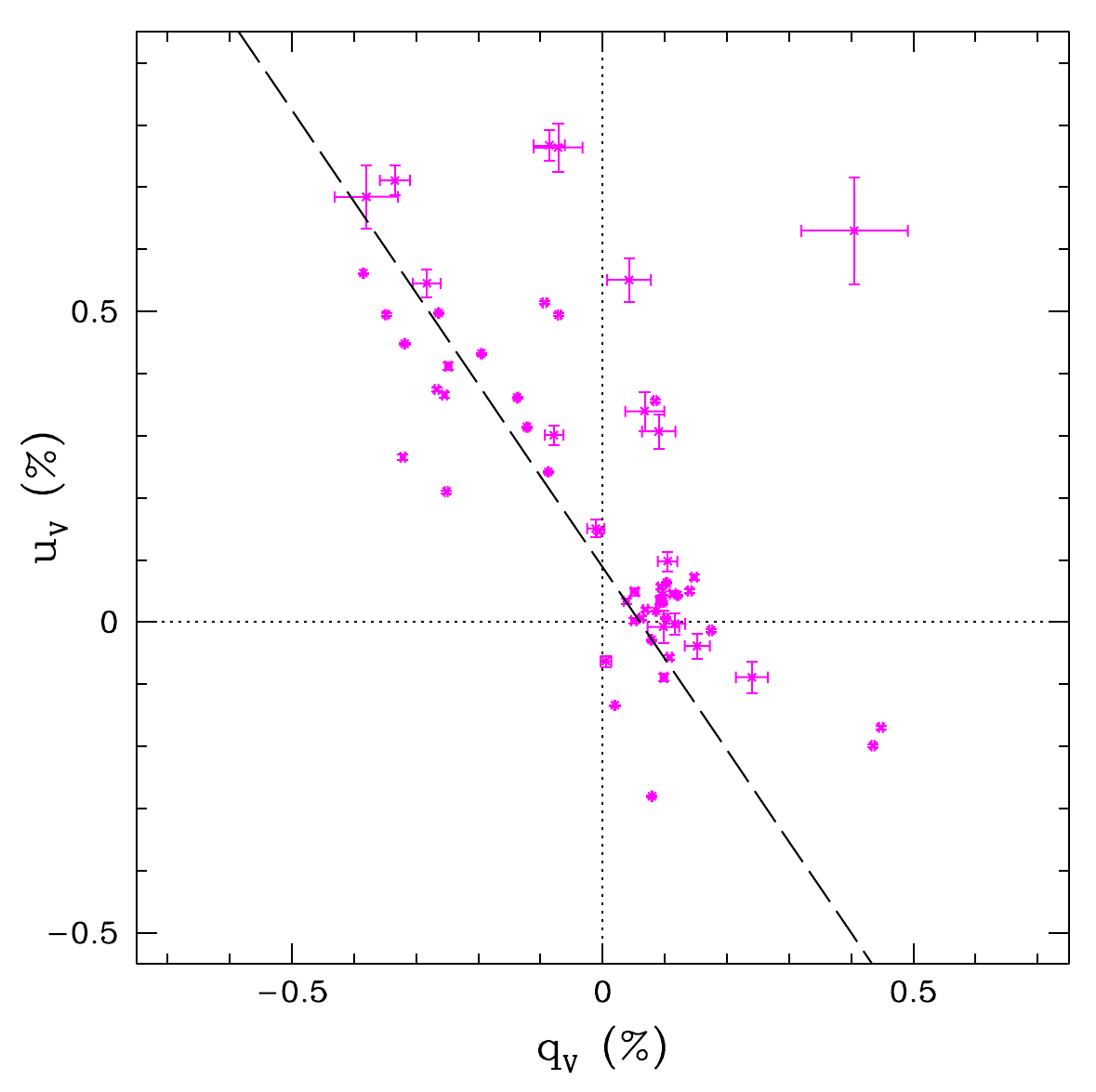}
\caption{An illustration of Category I polarimetric variability mainly along a line in the $q-u$ diagram. This dataset displays V-band photopolarimetry of Mira from HPOL in the years 1989-2004, \citep{2024AAS...24320719P}. The dashed line is a regression to the data. Although some points lie off the line fit, variations are reasonably defined by this axis for the majority of the data, indicating a relatively stable orientation of the system PA star's aspherical geometry.}
\label{figW}
\end{figure}

\begin{figure}[t]
\centering
\includegraphics[width=3.in]{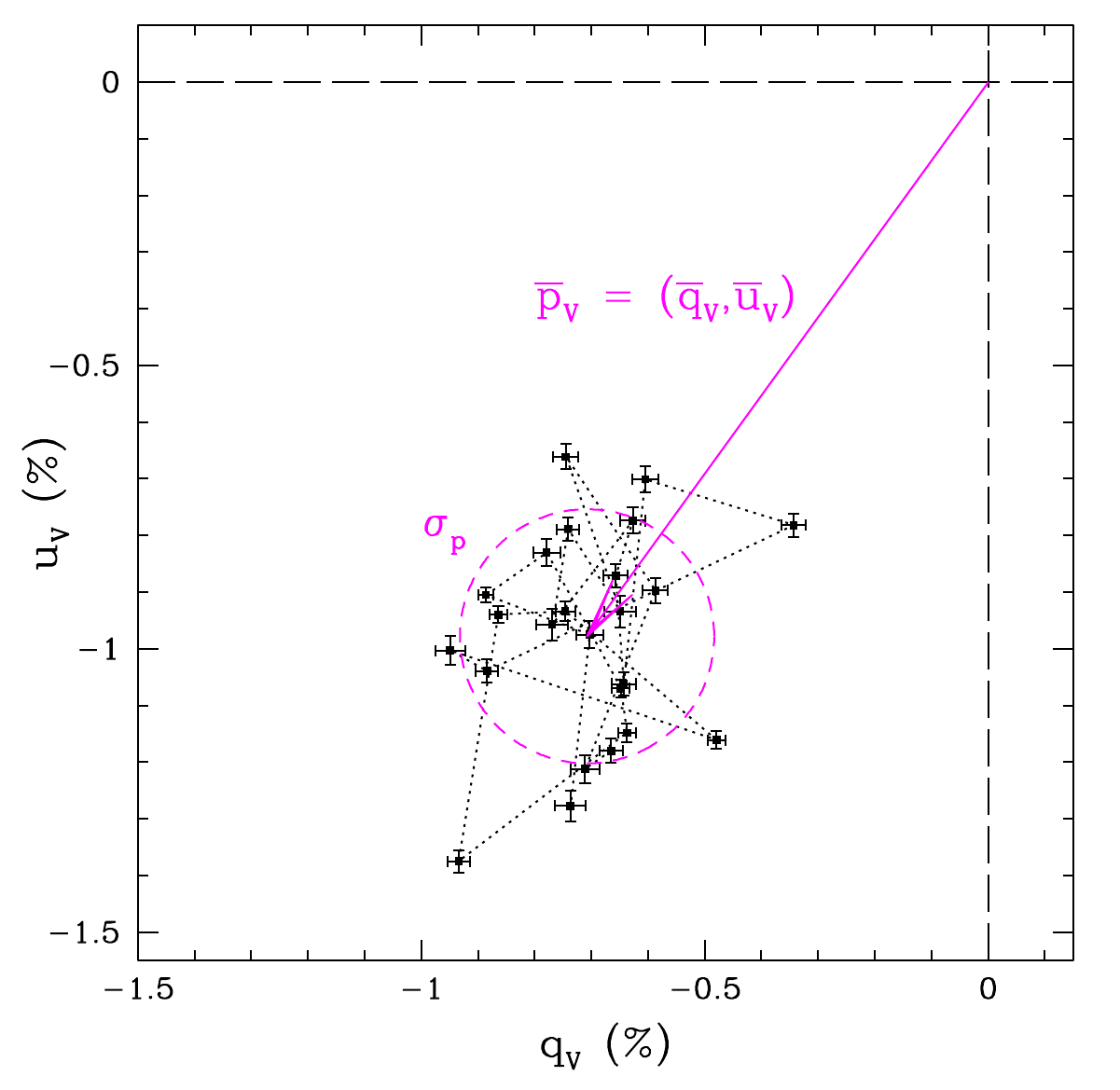}
\caption{An example of inferring ISP for the Category III for stochastic variable polarization. Data for the WN8 star WR~40 \citep{2023MNRAS.519.3271I} are plotted in the $q-u$ plane connected by dotted lines in time-sequential order. The error-weighted centroid is used to define the location for the ISP as identified by the pink vector. The dashed pink circle is labeled $\sigma_{\rm p}$ for the standard deviation of the measurements about the constant polarization, a spread that is related to the physical properties of the clumped wind.}
\label{figX}
\end{figure}

It is useful to review some observed examples of the categories I--III introduced in Figure~\ref{figV}, along with spectral-based approaches not covered by that figure.  Two for time-variable effects are given in Figures~\ref{figW} and \ref{figX}. The first of these shows HPOL\footnote{The data presented here were obtained between 1989 October 1 and 2004 September 10, when HPOL was located at the 36" telescope at Pine Bluff Observatory (University of Wisconsin-Madison).} observations of the semi-regular variable star Mira, as synthetic V-band photopolarimetry based on the spectropolarimetry. The V~photopolarimetry data are plotted in a $q-u$ diagram \citep{2024AAS...24320719P}. The dashed line is a linear regression to the dataset. While not all the data lie on or near this line, there is a clear indication of a preferred axis for geometry at Mira. Although the variations are intrinsic the star, it is not possible to remove the ISP without additional information.

The second example is the WN8 star WR~40, shown in Figure~\ref{figX}. These photopolarimetric data were obtained with the Minipol polarimeter when it was mounted on the 0.6-m telescope at Las Campanas, Chile in early 1990 \citep{2023MNRAS.519.3271I}. Black points are the measurements, which display a largely random distribution. The black dotted line connects the points in chronological order, with most obtained over a period of about six weeks. The centroid of the distribution is taken to be the ISP, with $\bar{p}_V = (\bar{q}_V,\bar{u}_V)$ shown as a pink arrow. Also displayed is a pink dashed circle labeled $\sigma_{\rm p}$ for the standard deviation of the polarization. It is this property of the data that would be used to model the stochastically structured wind flow of the star.  While the random distribution of points is consistent with a spherical wind in time-average, there could be a time-average intrinsic polarization to the star \citep[e.g.,][]{2020ApJ...900..162G}.

\begin{figure}[t]
\centering
\includegraphics[width=3in]{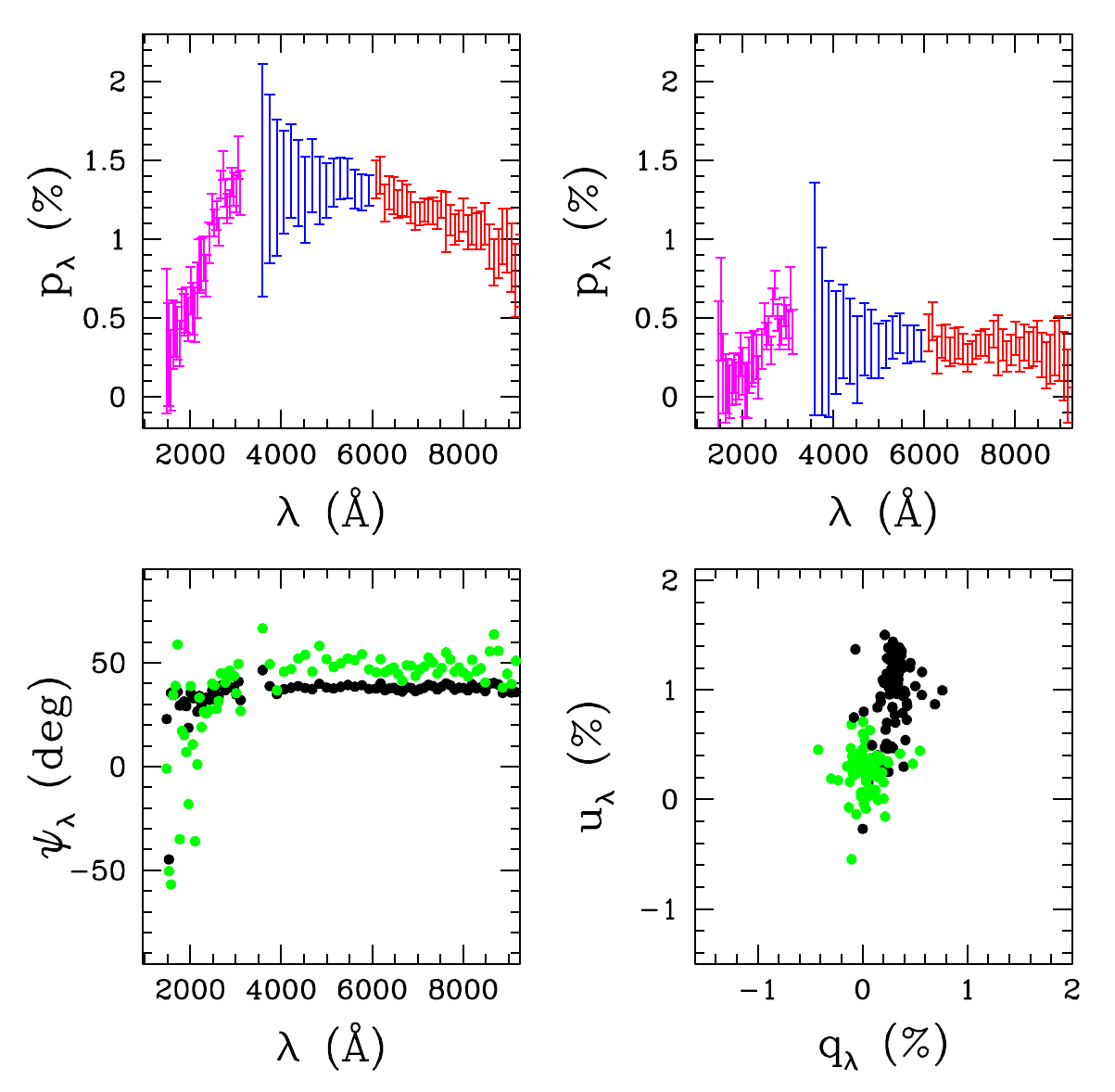}
\caption{An illustration of removing ISP for P~Cygni using data from WUPPE and HPOL \citep{2020ApJ...900..162G}. The UV data are shown in magenta. The ground-based data consists of two parts: shorter wavelengths in blue and longer ones in red. The data have been binned significantly to increase signal-to-noise. The upper left panel shows measurements with the ISP included. The upper right panel shows the result when ISP is removed under the constraint of achieving a mainly ``flat'' polarized spectrum as expected from electron scattering \citep{2020ApJ...900..162G}. The lower panels compare before and after correction as green and black, respectively, for the polarization angle (lower left) and in the $q-u$ diagram (lower right).}
\label{figY}
\end{figure}

\begin{figure}[t]
\centering
\includegraphics[width=3.in]{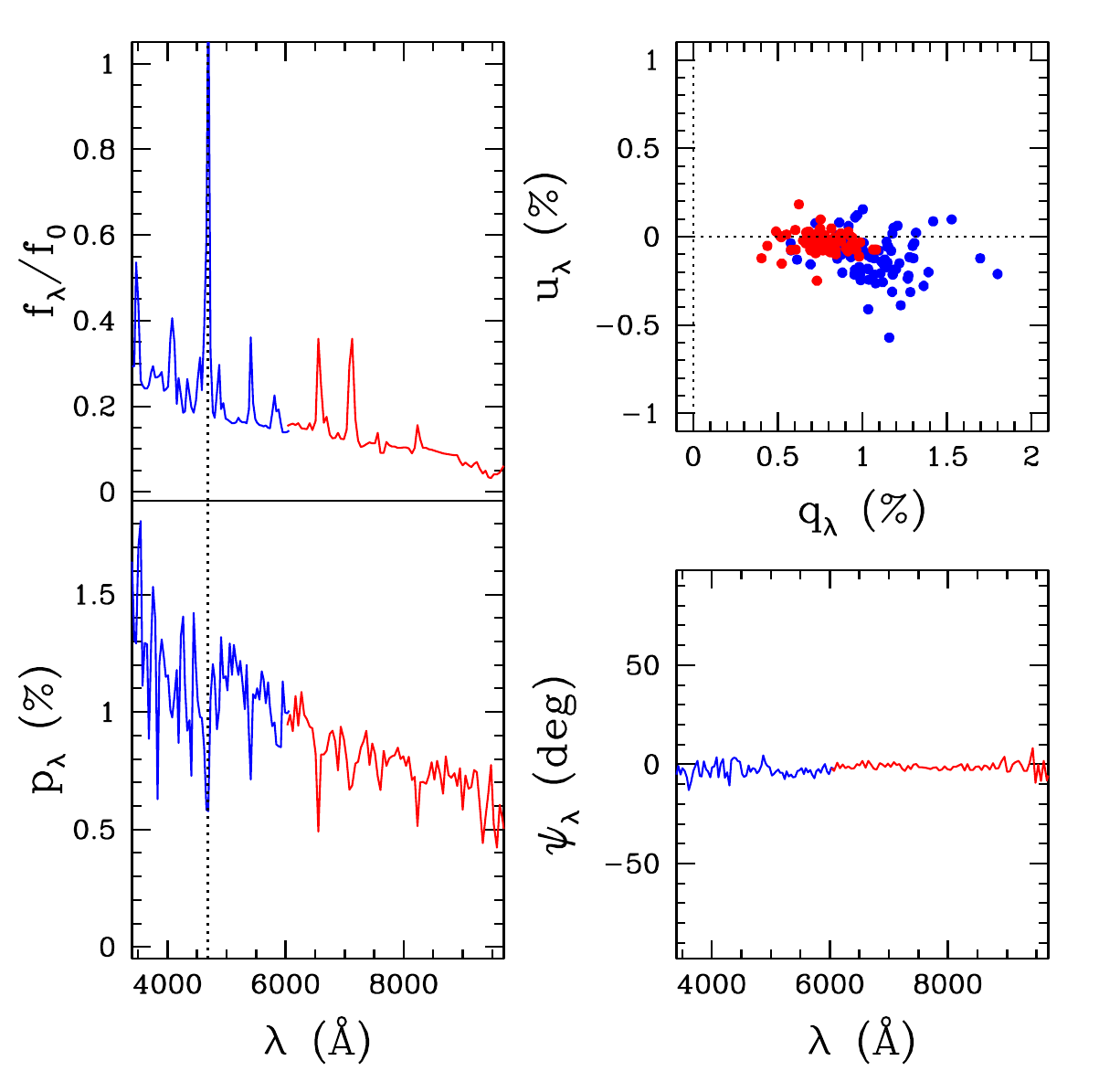}
\caption{Another valuable diagnostic to help disentangle the ISP from the intrinsic stellar polarization is the ``line effect.'' The upper left panel is the HPOL flux spectrum of the WN5 star WR~134, as in Figure~\ref{figX} with blue and red signifying the separate spectral bands. The strong emission line is He~{\sc ii} 4868, with a vertical dotted line as a visual guide. Lower left is the polarization spectrum showing depolarization across all of the strong emissions lines, but with He~{\sc ii} 4868 emphasized. The fact that polarization does not plummet to zero suggests ISP may be present. The upper right panel displays points for the polarized spectrum in the $q-u$ plane, whereas the lower right panel shows polarization PA. Note the PA is largely flat, even across the emission lines. This could imply an ISP of similar PA, or that line photons are scattered in the circumstellar environment.}
\label{figZ}
\end{figure}

Wavelength dependence in the polarization can in some cases be employed to infer or constrain the ISP. Here two examples are considered.  The first is shown in Figure~\ref{figY} for the luminous blue variable (LBV) star P~Cygni \citep{2020ApJ...900..162G}. The upper panels are for the polarization with wavelength (left: no correction for ISP; right: after removal of ISP). WUPPE2 data are shown in magenta, and HPOL data are in blue and red, representing the two spectral segments used in their measurements. The magenta, blue, and red data have been binned by differing amounts to achieve an improved signal-to-noise ratio. In this example the choice of ISP removal was motivated by the assumption that the polarized spectrum should be consistent with gray electron scattering.  This implies two things:  (1) the polarization is constant or ``flat'' with wavelength after removal of ISP and (2) the position angle is constant with wavelength after removal of ISP.  The authors varied the three parameters for the Serkowski Law until a most ``flat'' polarization distribution was achieved. The lower panels show the PA (as $\psi_\lambda$) with wavelength (left) and the spectrum plotted as dots in the $q-u$ diagram (right). There green points are before ISP removal and black points are after removal.

The last example is Figure~\ref{figZ}, showing spectropolarimetry of the WN5 star WR~134 obtained with HPOL \citep{1992ApJ...387..347S}. This also has four panels, with the upper left now containing the flux spectrum (normalized); the lower left is the polarization; the upper right is the $q-u$ diagram; and the lower right is the PA. The left panels show a vertical dotted line for the position of the He{\sc ii} 4686 booming stellar wind emission line. This line emission is formed over a large volume. The polarization in the lower left panel shows a significant drop across the spectral line. The conclusion here is that the line photons are minimally scattered, and because polarization is a difference over a total, the result is a dilution (by normalization) of the continuum polarization, often known as the ``line effect'' \citep{1990ApJ...365L..19S}. The fact that the polarization does not go to zero suggests there is ISP. In other words, the dilution only acts on the stellar polarization, not the ISP, which is imposed as starlight travels from the source to the Earth. This line effect is another way of determining or constraining parameters for the Serkowski Law. Although only the 4686~\AA\ line is being emphasized, there are numerous strong emission lines that show depolarizing dilution effects.

\section{Inferring Intrinsic Stellar Polarization}\label{sec:infer_intrinsic_pol}

For hot massive stars, the primary polarigenic opacity is electron scattering. From the UV and longward, electron scattering is well approximated by gray Thomson scattering. The major consequence of this is that linear polarization may naively be expected to be at a constant level and fixed PA with wavelength. However, the scenarios for gray versus chromatic effects in the polarization can be more diverse.

First, in situations where polarization arises entirely from scattering by circumstellar electrons, and the only opacity in the medium is electrons, and the only illumination by starlight, then indeed the expected polarization will be ``flat'' \citep{1977A&A....57..141B}. The medium will have a fixed scattering optical depth with wavelength, so the amount of scattered light is achromatic. Even if the system is spherically symmetric, there will be scattered light. If the system is aspherical, only then will a net polarization result, given by a ratio of the polarized flux $f_{\rm p}(\lambda)$ to the total flux $f_\ast(\lambda)+f_{\rm sc}(\lambda)$, which consists of direct starlight and scattered starlight. For this scenario, the polarized flux is a constant multiplied by the stellar flux. This is true even if the medium is optically thick to scattering. 

There are many situations in which chromatic effects arise for polarization for massive stars. The simplest is when there is more than one star \citep{1978A&A....68..415B}. Take again the scenario of only circumstellar scattering by electrons, and consider a binary star system in which the primary has a spherical envelope. Scattered light from the primary will produce no net polarization for the unresolved source. However, if the medium is spherical around the primary, that same envelope is clearly not spherical around the secondary, and symmetry is broken for that illuminating component. Nearly all massive stars are born into binaries, triples, or even higher order multiples \citep[e.g.,][and sources therein]{2011IAUS..272..474S,2013ARA&A..51..269D}. The occurrence of two illuminating sources with distinct spectral energy distributions (SEDs) will lead to wavelength-dependent polarization \citep[e.g.,][]{2022ApJ...933....5I}. 

Even in single stars (binaries with widely separated components), chromatic effects can appear owing to absorptive opacities, such as line blanketing \citep[e.g.,][]{1991ApJ...383L..67B}, or changes in polarization over individual strong emission lines \citep[e.g.,][]{1990ApJ...365L..19S}. Additionally, some massive stars are near-critical rotators. The stars are oblate -- distorted from spherical -- and can show gravity darkening and polar brightening. Such effects can lead to significant polarizations, especially at UV wavelengths \citep{1991ApJS...77..541C, 2017NatAs...1..690C}.

For our study of how intrinsic stellar polarization combines with an interstellar contribution, we consider four basic examples in the following sections.

\subsection{Polarization from Electron Scattering Only}

This is the simplest intrinsic polarization. We ignore spectral lines and absorptive opacity. Whatever the deviation from spherical, the intrinsic polarization is given by $p_\ast$ at all wavelengths and is therefore a flat polarized SED, regardless of the shape of the flux $f_\lambda$. We assign a stellar polarization PA on the sky, $\psi_\ast$, relative to an observer-defined system. The resultant Stokes parameters for linear polarization are

\begin{eqnarray}
q_\ast & = & p_\ast\,\cos 2\psi_\ast \\
u_\ast & = & p_\ast\,\sin 2\psi_\ast.
\end{eqnarray}

\noindent In the examples that follow, the intrinsic stellar polarization will always be axisymmetric, in order to illustrate its effects. One can always introduce an angle $\psi_\ast$ in relation to an observer's system, as described above.

Although no chromatic effects are considered in
this example, the polarization could be time dependent, through
$p_\ast(t)$ and/or $\psi_\ast(t)$. We will explicitly consider time-dependent behavior in context of binary stars.

\subsection{Rotational Distortion}

In this case, high rates of rotation lead to the geometrical distortion of the star with associated brightness variations. In the approach of \citet{1924MNRAS..84..665V}, $T_{\rm eff}^4 \propto g_{\rm eff}(\theta)$, for latitude $\theta$. A more recent treatment has been considered by \cite{2011A&A...533A..43E} (ELR). In either case, the distortion of the star is attended by polar brightening and equatorial darkening. In this example, we ignore any circumstellar material, and a polarized SED results entirely from the axisymmetric stellar atmosphere. This implies a fixed polarization PA; however, the effects of gravity darkening are inherently chromatic because of radiative transfer effects, generally with much higher polarizations toward the UV for hot massive star atmospheres. 

\begin{table*}[t]
\begin{center}
\centering
\caption{Model Parameters \label{tab1} }
\begin{tabular}{cccccccr}
\hline\hline Figure & \multicolumn{4}{c}{Interstellar Properties} & \multicolumn{3}{c}{Stellar Properties}\\
 & $p_{\rm max}$ & $\psi_I$ & $\lambda_{\rm max}$ & $K$ & $p_\ast$ & $\psi_\ast$ & Sec. \\ %$^a$ \\ 
 & (\%) & (deg) & ($\AA$) & & (\%) & (deg) & \\ \hline
\ref{fig7} & 0.5 & 30 & 5500 & 0.9 & 0.5 & 0--80 & \S 5.1 \\
\ref{fig8} & 0.5 & 30 & 5500 & 0.9 & ---$^a$ & 75 & \S 5.2\\
\ref{fig10}-\ref{fig11} & 0.5 & 30 & 5500 & 0.9 & ---$^b$ & 63 & \S 5.3 \\
\ref{fig12} & 0.5 & 30 & 5500 & 0.9 & ---$^b$ & 0$^c$ & \S 5.4 \\ \hline
\end{tabular}
\flushleft{\small
%\noindent $^a$ ``Kind'' refers to the cases as subsection numbers. \\
\noindent $^a$ Intrinsically chromatic and values depends on $\omega$; see text for details.\\
\noindent $^b$ Intrinsically chromatic. \\
\noindent $^c$ A non-zero value of $\psi_\ast$ serves only to rotate the elliptical loops by that value in the $q-u$ diagram.}
\end{center}
\end{table*}

\subsection{Optically Thick Winds}

This example allows for multiple scattering by electrons, plus absorption effects in the continuum (such as free-free and bound-free absorption), along with line blanketing. Additionally, changes across individual spectral lines are explored. For this example, a synthetic spectrum for an axisymmetric classical Wolf-Rayet (WR) wind of WNh subtype is used \citep{1996MNRAS.281..163S}, from a calculation with the CMFGEN code \citep{1998ApJ...496..407H}.

Our example is a wind with a latitudinal density profile given by

\begin{equation}
\rho(r,\theta) = \rho_{\rm pole} + \left(\rho_{\rm eq}
  - \rho_{\rm pole} \right)\,\sin^2 \theta,
\end{equation}

\noindent where $\rho_{\rm eq}/\rho_{\rm pole} = 3.3$, and the wind is otherwise radially streaming (i.e., the wind flow has
only a radial velocity component). At a Rosseland optical depth
of 20, the star has a temperature $T_\ast=48,000$~K, radius
$R_\ast = 18.9~R_\odot$, luminosity $L_\ast = 1.72\times 10^6~L_\odot$, with $\log g = 3.818$. The global mass-loss rate is $4.5\times 10^5~M_\odot/$yr with a terminal speed of $v_\infty = 2100$~km/s. The mass fraction abundances are $X=0.201$, $Y=0.796$, and $Z=0.003$.

\subsection{Binary System}

The last example is for a binary system which will generally show time-variable polarization that is cyclical on the orbital period of the system. Here we draw on the optically thin scattering treatment of \cite{1978A&A....68..415B} involving two point sources of illumination and a static circumstellar envelope in the co-orbiting frame. 

Chromatic effects arise because the two illuminating sources have different SEDs. In our treatment, we define the two stars as the primary and the secondary, with specific luminosities $L_{1, \lambda}$ and $L_{2, \lambda}$, respectively. For a scattering medium about the stars that is axisymmetric with respect to the line centers for the binary, contributions to the polarization from each separate star are $p_1$ and $p_2$ as amplitudes \citep{2022ApJ...933....5I}. In combination, the time- and wavelength-dependent polarization is given by

\begin{equation}
p_\lambda (t) = \frac{L_{1,\lambda}\,p_1(t)+L_{2,\lambda}\,p_2(t)}{L_{1,\lambda}+L_{2,\lambda}}.
\end{equation}

\noindent where time dependence is cyclic on the orbital period. The limiting behavior of the expression is that $p_\lambda$ becomes $p_1$ at wavelengths or times when the primary is dominant, and vice versa. Consequently, the polarization is flat over ranges of wavelength where either the primary or secondary act as the dominant source of illumination. 

\section{Hybrid Polarized SEDs}\label{sec:hybrid_seds}

To understand how ISP as expressed by the Serkowski Law can influence the net observed polarization, we introduce a PA for the interstellar contribution as $\psi_I$. The wavelength-dependent ISP is $p_I(\lambda)$ as given by the Serkowski Law in equation~(\ref{eq:serk}). The corresponding Stokes parameters are

\begin{eqnarray}
q_I(\lambda) & = & p_I(\lambda) \,\cos 2 \psi_I \\
u_I(\lambda) & = & p_I(\lambda) \, \sin 2 \psi_I.
\end{eqnarray}

Although the Serkowski Law introduces a chromatic signal to be combined with the stellar one, its form is known, even if the parameters $p_{\rm max}$, $\lambda_{\rm max}$, and $K$ may not be. In particular, there are three key characteristics that can be assumed for ISP:

\begin{enumerate}

\item The polarization peaks in the optical and drops toward both shorter and longer wavelengths, such that the ISP becomes less confounding for stellar signals in the UV and IR bands. 

\item Although chromatic, ISP is time-independent. Consequently, all time-variable polarization is intrinsic to stellar sources. While one may not be able to infer a constant component to the stellar polarization, all variable polarization is entirely stellar in origin. 

\item ISP is a smoothly varying function with wavelength, not showing any narrow or abrupt spectral features\footnote{In the UV, the only ISP feature is the 2175~\AA~bump, which is rarely polarized \citep{1997ApJ...478..395W}.}. Changes of polarization over spectral lines from the star, or other features such as bound-free edges, imply the star has an intrinsic polarization. The Serkowski Law is normalized to the input total flux. If the star were unpolarized, no matter how complex its spectrum, the Serkowski Law would result upon normalization. The variability over sharp spectral features thus implies a contribution to the polarization that is intrinsic to the star.

\end{enumerate}

\begin{figure}[t]
\centering
\includegraphics[width=3.in]{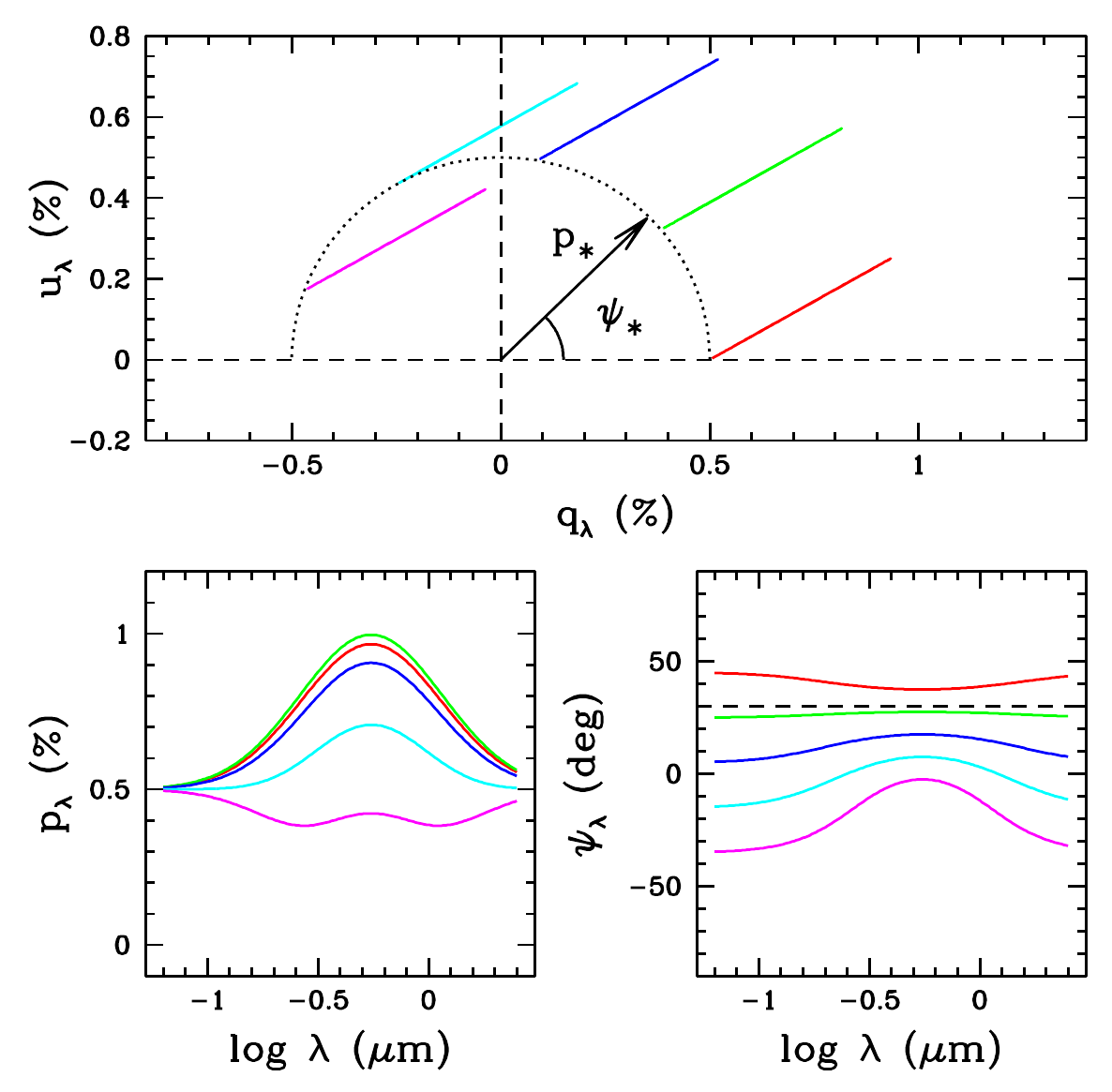}
\caption{Illustration for how ISP and intrinsic source polarization combine when the polarigenic opacity for the source is only Thomson scattering. The polarization of the star is fixed at $p_\ast=0.5\%$, but with five different values of $\psi_\ast$ for the five different colors. We adopt an ISP with $p_{\rm max} = 0.5\%$ and $\psi_I=30^\circ$.  The latter is the horizontal dashed black line in the lower right panel. Refer to the text for more description.}
\label{fig7}
\end{figure}

\begin{figure*}[t]
\centering
\includegraphics[width=5in]{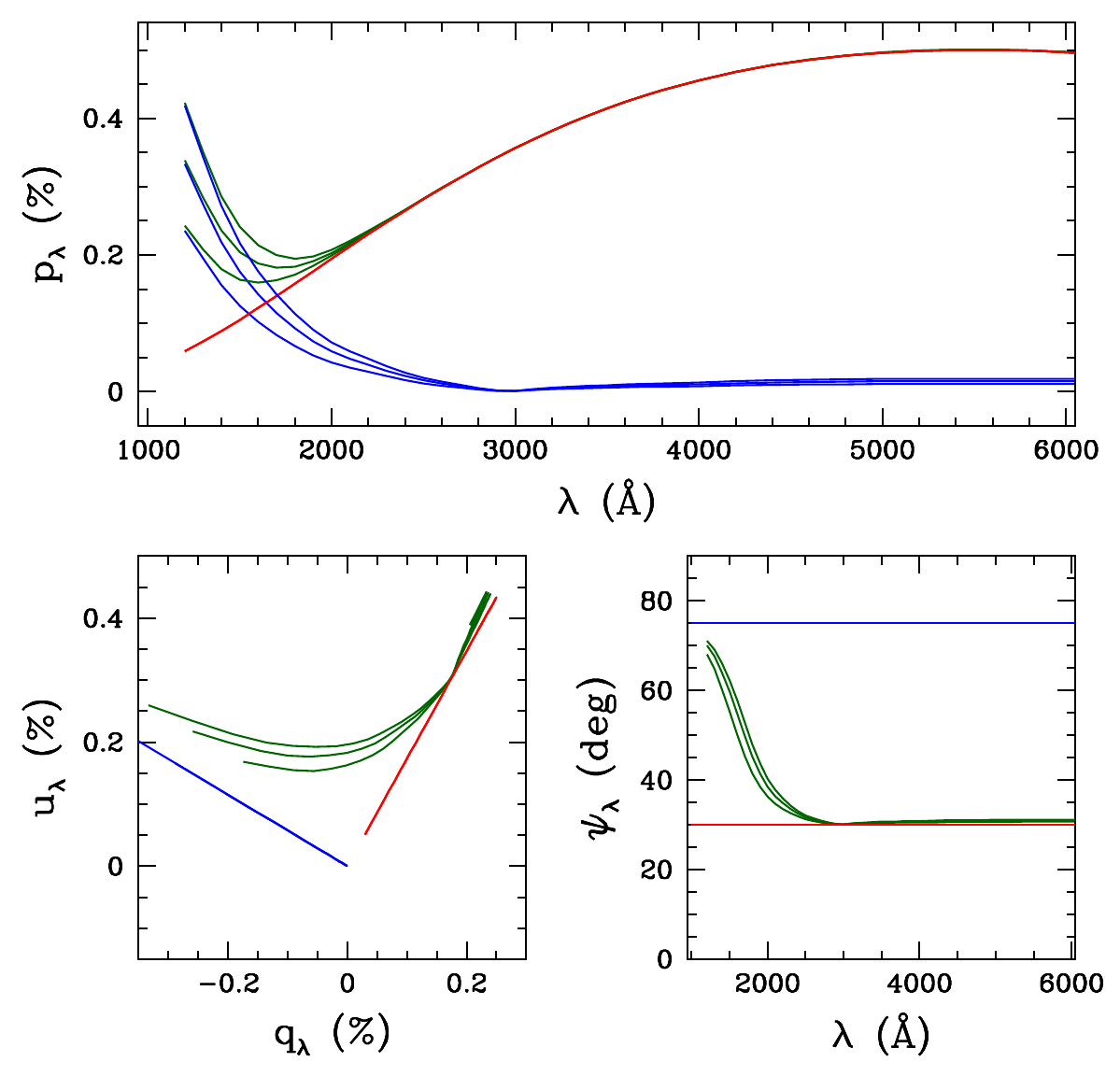}
\caption{This example illustrates how ISP combines with a source showing influences from rotational distortion.
Here, the format is similar to Figure~\ref{fig7}, but with $q-u$ in the lower left panel and the polarized SED at the top. Blue is the intrinsic stellar polarization; red is for a Serkowski Law with in Tab.~\ref{tab1}; and green is the combined polarization. The three blue curves are for models of a B1~V star with near-critical rotation rates of $\omega=$~0.9, 0.95, and 0.975. The models use von Zeipel gravity darkening and assume the star are viewed equator-on.}
\label{fig8}
\end{figure*}

With the Serkowski Law and above characteristics, creating ``hybrid'' polarized spectra is relatively straightforward. The Stokes parameters for the stellar and interstellar components add linearly as

\begin{eqnarray}
q_\lambda (t) & = & q_I(\lambda) + q_\ast(\lambda,t) \\
u_\lambda(t) & = & u_I(\lambda) + u_\ast(\lambda,t).
\end{eqnarray}

\noindent Then the total polarization and associated net PA are

\begin{eqnarray}
p_\lambda(t) & = & \sqrt{q^2+u^2} \\
\tan 2\psi_\lambda(t) & = & u/q = \frac{u_I(\lambda) + u_\ast(\lambda,t)}{q_I(\lambda) + q_\ast(\lambda,t)}.
\end{eqnarray}

\noindent There are in fact cross-terms \citep{2010A&A...510A.108P}. However, both the stellar and the interstellar polarizations are small, so cross-terms are ignored in our treatment. 

The following sections present results for hybrid polarized SEDs corresponding to the stellar polarization cases of the previous sections. Selected parameters are summarized in Table~\ref{tab1} as case-by-case (last column as ``Kind'') and corresponding figures. Note the absence of entries in the column for $p_\ast$ for stellar polarizations that are intrinsically chromatic, since no one value represents those cases.

\subsection{Hybrid with Thomson Scattering Only}

Figure~\ref{fig7} shows the case of combining a ``flat'' electron scattering SED with a Serkowski Law.  The upper panel is a $q-u$ diagram with lower panels showing how polarization (left) and PA (right) vary with wavelength.  The invariant in this example is the value of $p_\ast$ for the intrinsic polarization.  As an example, $p_\ast = 0.5\%$ is chosen.  The dotted circle is for this fixed intrinsic stellar polarization as a function of $\psi_\ast$.

The five colored curves signify how the observed polarization is influenced by an interstellar contribution.  For the Serkowski Law,  values of $p_{\rm max} = 0.5\%$, $\lambda_{\rm max} = 0.55~\mu$m, $K=0.9$, and $\psi_I=30^\circ$ are fixed.  The five colored curves are for values of $\psi_\ast = 0^\circ, 20^\circ, 40^\circ, 60^\circ,$ and $80^\circ$. 

In the upper panel, the inclusion of ISP leads to linear excursions that stem from the circle at each respective value of $\psi_\ast$.  The direction of each linear excursion is set by the fixed $\psi_I$.  Changing $\psi_I$ would rotate the excursions.  Recall that the Serkowski law is double-valued, so that in terms of wavelength, the straight lines are the chromatic variation running from zero ISP, where each excursion touches the circle, out to maximum value set by $p_{\rm max}$, then returning back to the circle.  This variation is how the Serkowski Law from UV through IR maps into the $q-u$ diagram.  If it were not for the stellar polarization, all five of these colored lines would degenerate to a single line hinged at the zero point of the plot with orientation set by the interstellar PA.

Hybrid polarized SEDs result from vector additions in the $q-u$ diagram at each wavelength to combine stellar and interstellar contributions. The bottom panels are more traditional ways to view chromatic effects in the polarization and PA. Whereas the $q-u$ diagram highlights the vector nature of the hybrid polarized SEDs, the lower left panel for total polarization traces the vector magnitude with wavelength. The lower right panel traces the variation in net PA. Clearly the stellar PAs are obtained at the shortest and longest wavelengths. In the optical, where ISP peaks, all curves pinch towards the interstellar PA, $\psi_I$.

\subsection{Hybrid with Rotational Distortion}

Figure~\ref{fig8} shows examples for a near-critical rotation star at $\omega=0.9$, 0.95, and 0.975, with higher $\omega$ corresponding to larger peak polarizations at the shortest wavelength. Gravity darkening shows a chromatic trend that is distinct from that of Serkowski: where the ISP is dropping toward UV wavelengths, the gravity darkening is rapidly becoming pronounced. Here, the panels are the same as Figure~\ref{fig7}, but are arranged differently: the $q-u$ diagram is shown in the lower left panel, and $p_\lambda$ is at the top. Each panel shows the intrinsic stellar polarization as blue, the ISP as red, and the total polarization as green. Note especially the lower right panel for PA, where the black curve transitions from the interstellar PA at longer wavelength (the horizontal red line) to that of the star (horizontal blue line).

\begin{figure}[t]
\centering
\includegraphics[width=3.in]{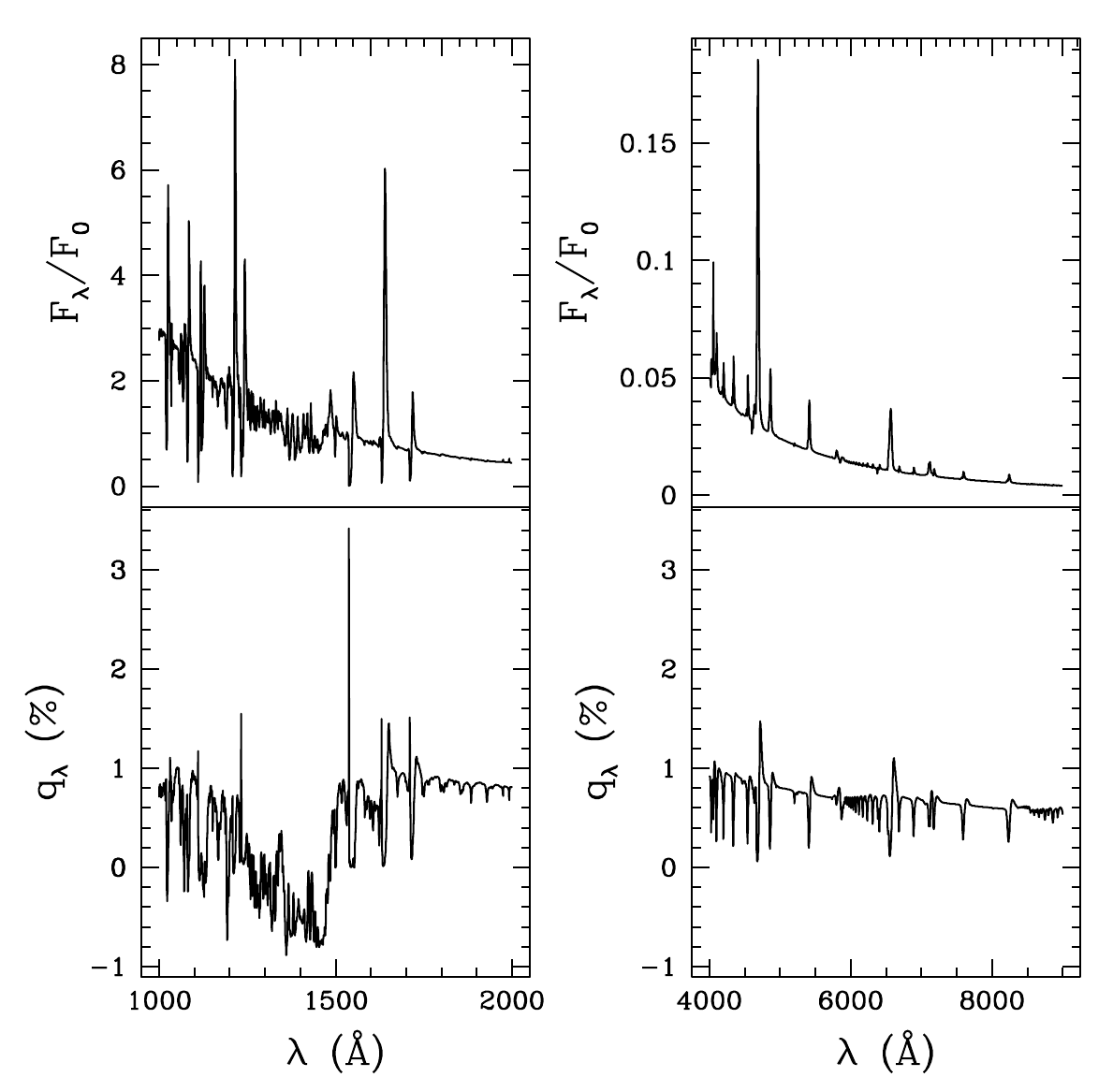}
\caption{An illustrative example of a dense circumstellar scattering medium showing intrinsically complex spectropolarimetric behavior. In this case, the calculation is for a WNh star. The upper panels show the UV (left) and visible (right) portions of the spectrum with $F_0 = 10^{14}$ erg/s/cm$^2$/\AA.} The lower panels shows the linear polarization. The wind is axisymmetric with an equatorial density that is $3.3\times$ denser than along the pole. Note the significant depolarization across the emission lines, although sometimes enhanced polarization is seen. The change of sign in the UV owes to iron-line blanketing, and signifies a polarization PA rotation by $90^\circ$, because more scattered light escapes from the polar wind. See the text for more details.
\label{fig9}
\end{figure}

For hot massive stars, the rotational distortion and associated atmosphere effects for temperature distributions show that the polarization is steeply rising in the UV. This is where the interstellar contribution is dropping. Any rising polarization cannot be interstellar in nature. The ability to recover how close the star is to critical rotation will depend on several factors. One of those is the specifics of the ISP. 

%The value of $p_{\rm max} = 0.4\%$ adopted here was chosen as an intermediate of the well-studied Be stars $\gamma$ Cas, $\phi$ Per, and $\psi$ Per at 0.26, 0.4, and 0.82\%, respectively \citep{1997ApJ...479..477Q}. 

The other consideration is the star itself. The example of Figure~\ref{fig8} for a B1V star is fairly conservative. Higher polarizations are expected of subgiant and giant stars where sphericity effects become more important \citep{1971MNRAS.154....9C, 2016A&A...586A..87K}. In our example, there is a fairly quick switchover (in wavelength) from the ISM-dominated to the star-dominated polarization, right around 1500~\AA. Shortward, the ISP hardly matters, so that in this case estimates of $\omega$ would be fairly accurate by fitting models shortward of the switchover. Indeed, the curvature of the switchover would help to characterize parameters for the Serkowski Law. 

\subsection{Hybrid with Optically Thick Winds}
\label{sub:thick}

Our third example involves an optically thick circumstellar medium with multiple scattering. For this purpose, we use an axisymmetric WNh wind model with total flux and polarized spectra provided by CMFGEN, with stellar and wind properties that were given in \S~4.3. The synthetic spectra in the UV and visible are displayed in Figure~\ref{fig9}. Its complexity is worth several remarks before a hybrid polarized SED. The upper panel is the total flux distribution, showing an approximately power-law continuum with numerous emission lines superposed. The two strongest lines at optical wavelengths are He{\sc ii} 4686~\AA~and H$\alpha$. The UV spectrum has several prominent emission lines and resonance lines, and notably a region of line blanketing by numerous iron lines around 1400~\AA.

In the lower panel for polarization, there are three main features. First there is a somewhat constant continuum polarization level, although it appears to be declining toward the longer wavelengths. This is expected because the thick pseudo-photosphere that forms in the wind itself becomes larger with increasing wavelength, which acts to absorb the inner scattered light by electrons. 

Second, there are polarization drops across the majority of the emission lines. This is the so-called dilution effect. Much of the line emission forms over a region that is extended compared to where the electron scattering forms the net continuum polarization. Since polarization is a relative quantity normalized by the total flux, strong emission lines increase the total flux without a proportionate increase of polarized flux. Thus in ratio, the polarization is diminished or diluted by virtue of normalization.

\begin{figure}[ht]
\centering
\includegraphics[width=3.in]{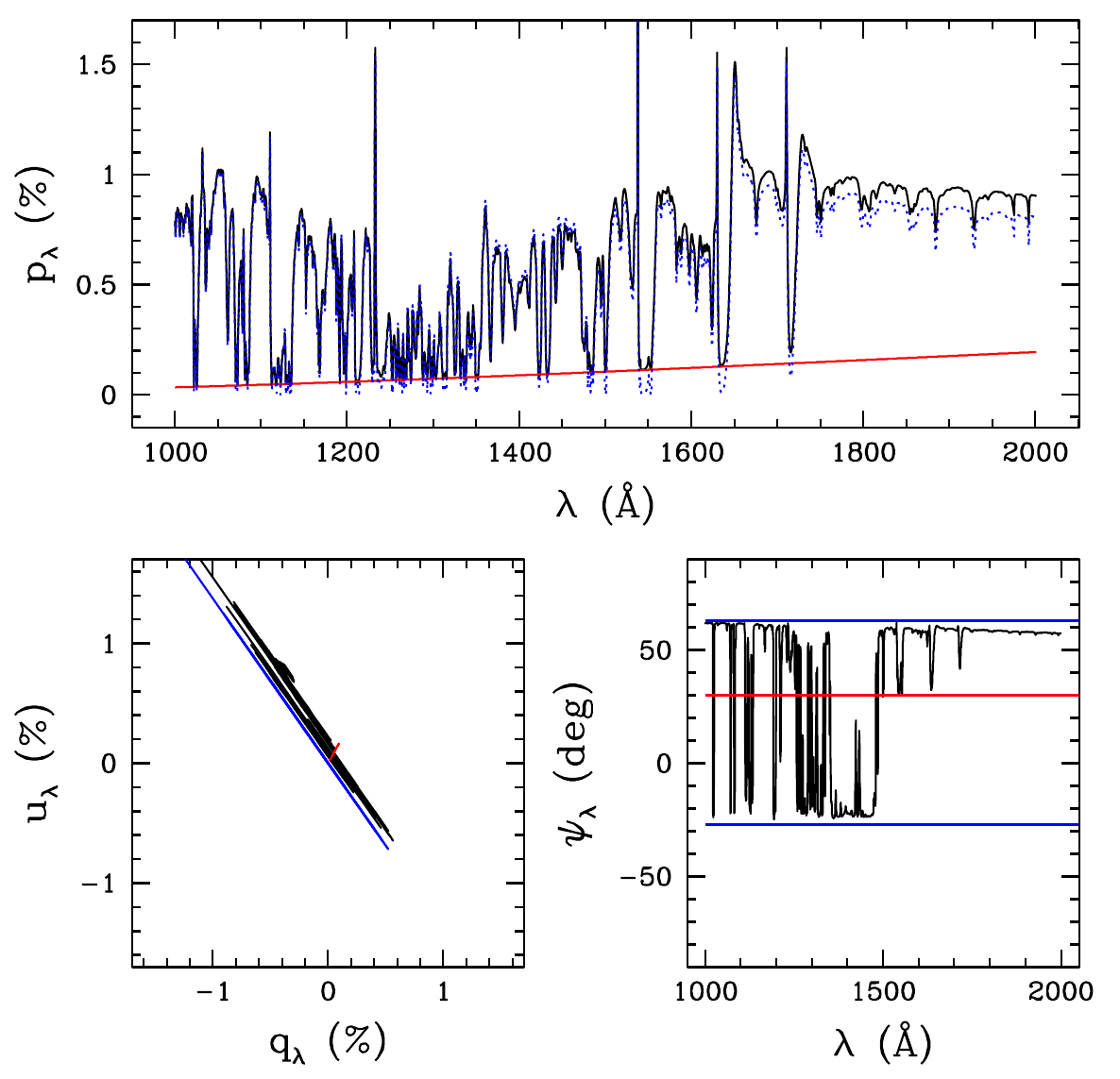}
\caption{A Serkowski Law with $p_{\rm max}=0.5\%$ is combined with the WNh model calculation from Figure~\ref{fig9}, using the same format as Figure~\ref{fig9}, but for the UV portion only. In the top panel, the intrinsic source polarization is shown in dotted blue lines (instead of solid blue) for clarity. For the lower right panel, the PA rotation indicates there are two limiting intrinsic source PAs, as indicated in blue.}
\label{fig10}
\end{figure}

\begin{figure}[ht]
\centering
\includegraphics[width=3.in]{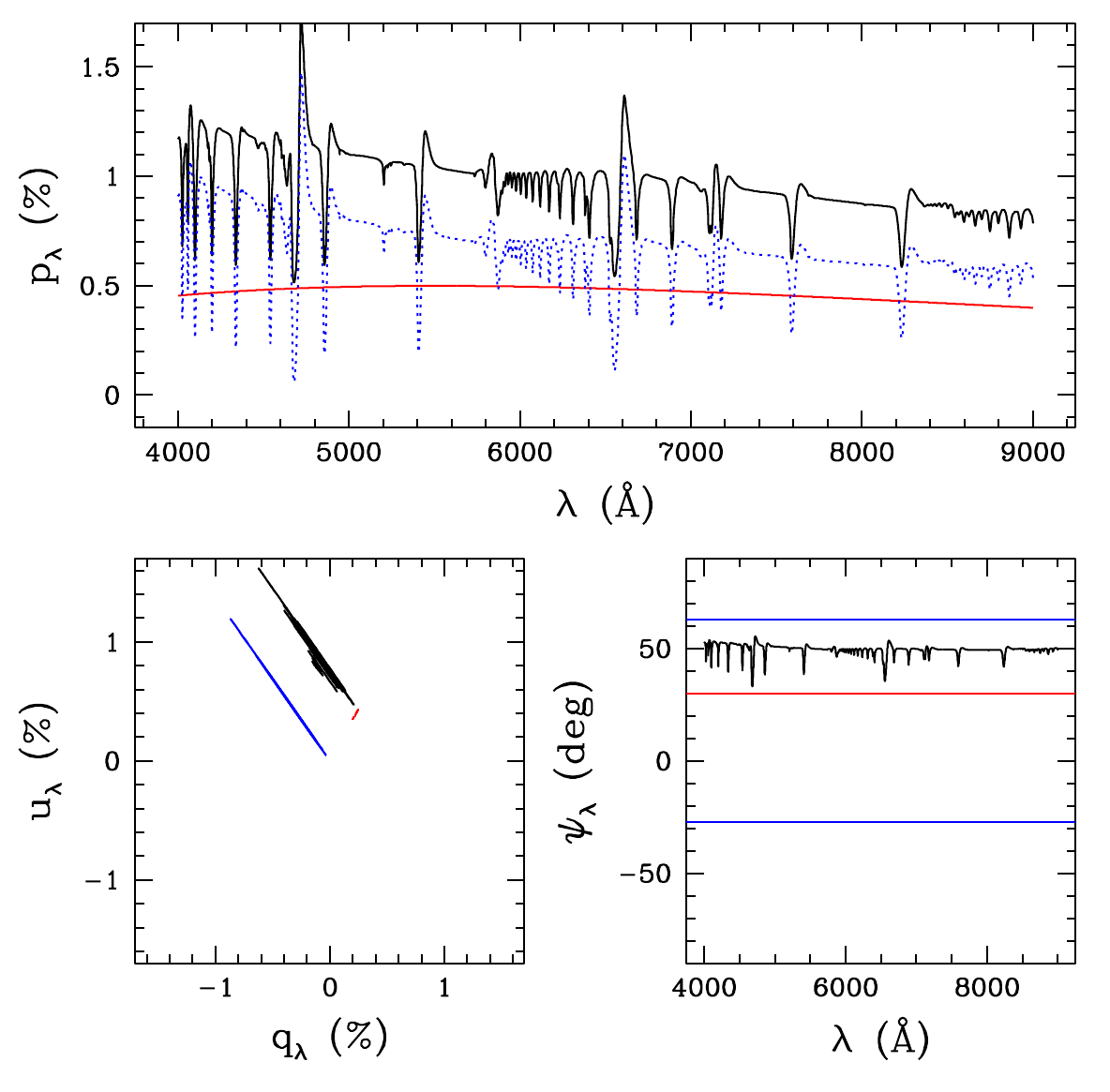}
\caption{Same style as Figure~\ref{fig10}, now emphasizing the visible band. }
\label{fig11}
\end{figure}

Third, across the forest of iron lines, the polarization actually switches signs. The equatorial density is $3.3\times$ larger than at the pole. The iron-line blanketing is more severe at equatorial latitudes, and so the surviving polarized flux at those wavelengths emerges from polar wind flow. Consequently, that polarization is rotated in PA by 90 degrees from the continuum outside the line blanketing. This is represented as a sign change in the polarization. Thus, in this example, through polarization, one can infer information about the distinct polar and equatorial zones over wavelengths demarcated by the PA rotation.

Figures \ref{fig10} and \ref{fig11} include the Serkowski Law for the UV and visible spectra, respectively, shown separately for clarity owing to the large number of rich spectral features. In both cases, blue is for stellar alone, red is for interstellar alone, and black is for the hybrid. For Figure~\ref{fig10}, the stellar polarization so dominates the ISP that the hybrid is shown as a dotted blue line for better clarity. Note in the lower right panel for PA that there are two blue horizontal lines highlighting the sign change in the model stellar polarization. Note also that the dilution effect in strong lines drops the hybrid polarization to the red curve level in the upper panel, and correspondingly identifies the interstellar PA in the lower right panel.

In Figure~\ref{fig11}, the situation is largely reversed in comparison with Figure~\ref{fig10}. At visual wavelengths, the ISP is stronger than stellar polarization for the parameters chosen here. The dilution effect across strong emission lines often -- but not always -- gets close to the red curve for ISP alone. The diminishing interstellar contribution and the presence of iron-line blanketing indicates that the UV is much better for inferring the intrinsic stellar polarization.

\subsection{Hybrid with a Binary System}

The final example is for a binary system. For illustration, we choose a binary with a circular orbit involving a hot primary at 28000~K and a cooler secondary at 18000~K. The primary has a radius of $5~R_\odot$ and the secondary of $18~R_\odot$. The parameters approximate the system of $\beta$~Lyr. Simple blackbody spectra are used to simulate chromatic effects.  Our choice of parameters means that the secondary is the more luminous star bolometrically, and that polarization arises from light scattered by the secondary only.  This means the light of the primary serves only to dilute the polarization.  Despite electron scattering being gray, the polarization becomes wavelength-dependent when there are two sources of radiation with different spectral energy distributions, as detailed by \cite{1978A&A....68..415B}.

\begin{figure}[t]
\centering
\includegraphics[width=3.in]{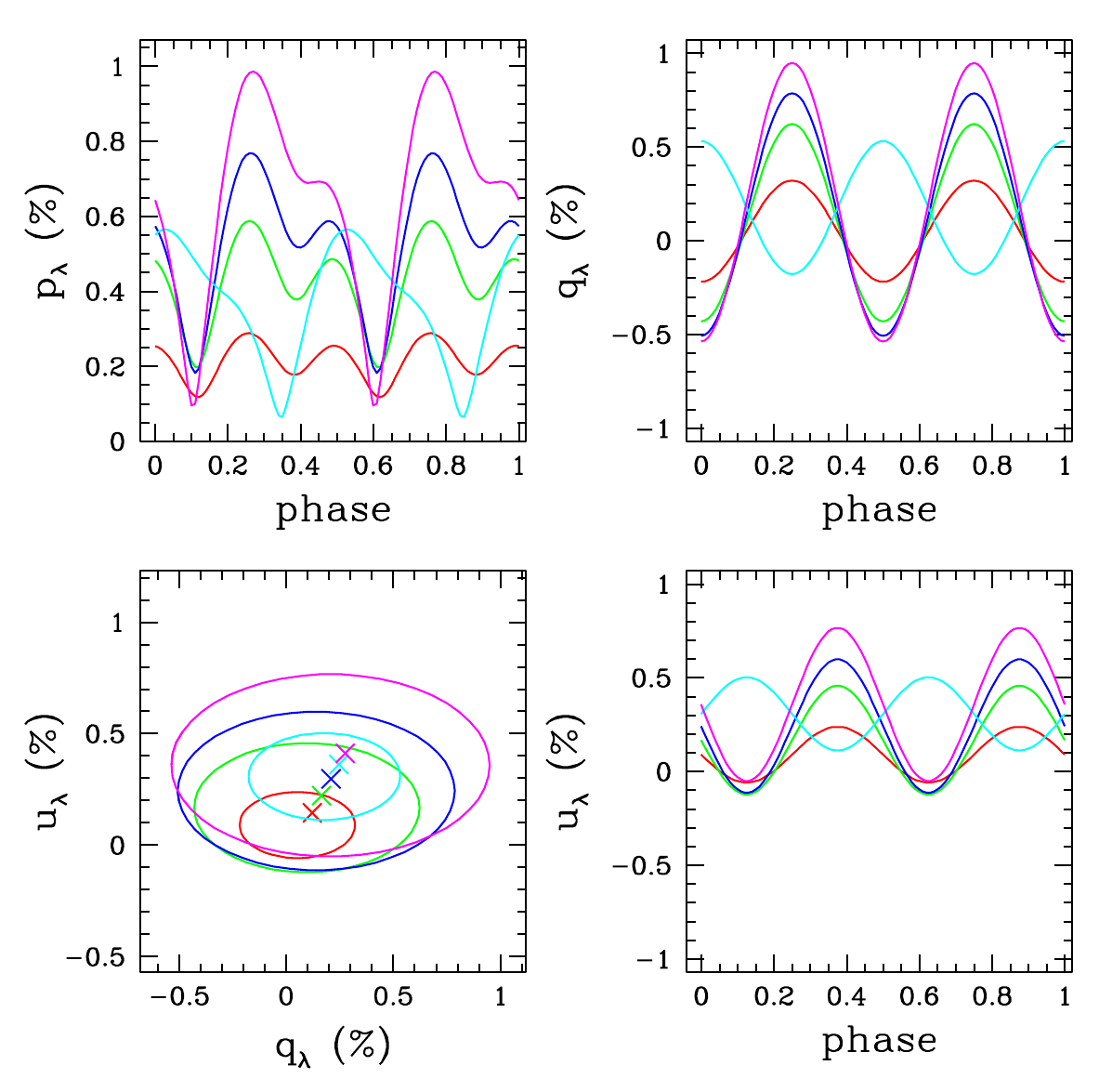}
\caption{An example involving variable polarization, a \cite{1978A&A....68..415B} model is used to represent an interacting binary consisting of a hotter primary with axisymmetric circumstellar disk and a cooler secondary (see the text). The system is viewed with an inclination of $i=72^\circ$.
Each curve is for a different wavelength from 1000-3000~\AA\ (red-1000; green-1500; cyan-2000; blue-2500; magenta-3000). The cyan curve at 2000~\AA\ has been artificially reduced and given a sign change (i.e., PA rotation of $90^\circ$) as can arise from line blanketing. Polarization parameters are plotted against orbital phase in the two upper panels and the lower right. Lower left is for a $q-u$ diagram where the small crosses indicate the ISPs (like those in Fig.~\ref{figV}) at corresponding wavelengths, and the elliptical loops are the intrinsic variable polarization of the source.}
\label{fig12}
\end{figure}

Our polarization model assumes an axisymmetric disk structure around the primary star with scattered light by the secondary and viewed at an inclination angle of $i=72^\circ$. Polarization light curves with orbital phase are calculated at five wavelengths of $\lambda=1000$~\AA\ to 3000~\AA\ in 500~\AA\ increments, as shown in Figure~\ref{fig12}. We artificially impose a sign change and 50\% reduction in the polarization at 2000~\AA\ to simulate iron-line blanketing. Such a sign change and consequent PA rotation is actually seen in the UV for $\beta$~Lyr \citep{1998AJ....115.1576H}; iron line blanketing can also suppress the UV polarization in Be decretion disks \citep{1991ApJ...383L..67B}.

In Figure~\ref{fig12}, the four panels are: total polarization at the upper left; Stokes $q_\lambda$ at the upper right; Stokes $u_\lambda$ at the lower right; and a $q-u$ diagram at the lower left. For the $q_\lambda$ and the $u_\lambda$ light curves with orbital phase, the cyan curve for $2000~\AA4$ with the imposed position flip stands out as being out of phase with variations for the other wavelengths. Note that while significant metal-line blanketing occurred around 1400~\AA\ in the model of the WNh wind in Section~\ref{sub:thick}, the effects is temperature dependent \citep[e.g., see Fig.~4 of ][]{2020Galax...8...60H}. We chose 2000~\AA\ as appropriate for a cooler and more common massive star.

The lower left panel is a $q-u$ diagram. Each curve is colored according to wavelength.  If viewed pole-on, the curves would all be circles for a circular orbit.  Away from pole-on, the curves are ellipses in the $q-u$ diagram, becoming increasingly eccentric with higher viewing inclination, degenerating finally to straight lines with no variation in $u$ for an edge-on view of the orbit with $i=90^\circ$.  Each curve is traced twice per orbit in the $q-u$ plane, corresponding to the second harmonic nature of the variations seen in the separate $q_\lambda$ and $u_\lambda$ lightcurves in the right panels. ISP values (see Tab.~\ref{tab1}) are indicated by the crosses of corresponding colors.  As indicated in Figure~\ref{figV} for Category ii, ISP serves only to translate the time-variable pattern in the $q-u$ diagram, not to alter its shape or its orientation.

\section{Conclusion}\label{sec:conclusions}

Spectropolarimetry provides important diagnostics of astrophysical sources through direct measures of magnetism (using circular polarization) and of geometry (using linear polarization). For linear polarimetry, a major challenge is that the ISM imposes a polarization onto the stellar value. For intrinsically unpolarized stars, that signal provides an opportunity to study the gas and dust of the ISM. When attention is centered on the intrinsic stellar polarization, the interstellar contribution can complicate the analysis.

We have explored several scenarios involving time- and wavelength-dependent stellar polarization as combined with an interstellar contribution. While the most ideal situation is to remove the ISP, this may not always be feasible, or may induce additional uncertainties in the measurements. We thus explored strategies to infer properties about the star even when the ISP is not removed.

Studies indicate that the Serkowski Law is a generally reliable prescription of ISP, indicating that from the optical, its influence declines toward the UV and the IR. This highlights the diminishing confounding impact of ISP as more polarimeters are developed for the IR and UV bands, such as the {\em Polstar} mission concept \citep{2024arXiv240915714I}.

Four illustrative stellar cases were considered: flat polarization from Thomson scattering, chromatic effects in continuum polarization from rotationally distorted stars, complex radiative transfer effects for thick winds, and both the chromatic and time variable polarization for a binary system. The chief conclusions are as follows.

\begin{itemize}

\item The case of Thomson scattering in optically thin circumstellar media highlights the advantages of conducting spectropolarimetry beyond the visible band. The contribution from the ISP declines toward both the UV and IR bands.  If the stellar contribution is expected to be gray from strictly thin electron scattering, multi-wavelength data can be used to infer the distribution of the ISP.

\item Rotational distortion highlights the value of UV spectropolarimetry for studying near-critically rotating hot massive stars. Radiative transfer in hot-star atmospheres with electron scattering leads to sharply rising polarization at FUV wavelength where the ISP is declining.  Instead of correcting for the ISP, its contribution may be negligible in comparison to the stellar signal.

\item The thick-wind WR example emphasizes the chance for depolarization across strong emission lines, but also the opportunities afforded by line blanketing at UV wavelengths, where polarization may trace different zones due to the effects of differential absorption in aspherical geometries. 

\item Finally, binarity provides an example of time-variable polarization that is entirely due to stellar contributions, with chromatic effects arising when two sources have differing SEDs.

\end{itemize}

It is clear that extracting information about the star from polarization when an interstellar contribution is present cannot be reduced to a single all-encompassing strategy. The approach depends on multiple considerations: (a) Does the polarization PA change with wavelength? (b) Does the polarization amplitude and/or PA change with time? (c) Does the polarization and/or PA show change over spectral features such as lines, edges, or broader regions from line blanketing? These attributes, when present separately or in combination, provide a means for separating the desired stellar signal from the ISP. 

\section*{Acknowledgments}

Comments provided by an anonymous referee have greatly
improved the presentation and clarity of the manuscript.
RI and CE gratefully acknowledge support from the National Science Foundation under grant number AST-2009412. DJH acknowledges partial support from the STScI through grant HST-AR-16131.001-A.
We would like to thank the HPOL project for their help in creating the MAST HPOL archive, especially Marilyn Meade from the University of Wisconsin who was our main contact with the HPOL and WUPPE project for the past several years.

\bibliography{ism_bib}

\end{document}